1# Stark Tuning of Donor Electron Spins in Silicon

Forrest R. Bradbury[1], Alexei M. Tyryshkin[1], Guillaume Sabouret[1], Jeff Bokor[2,3], Thomas Schenkel[2], Stephen A. Lyon[1]

[1]Department of Electrical Engineering, Princeton University, Princeton, NJ 08544
[2]E. O. Lawrence Berkeley National Laboratory, Berkeley, CA 94720
[3]Department of Electrical Engineering and Computer Science, University of California Berkeley, CA 94720

Corresponding author: bradbury@princeton.eduPACS: 71.70.Ej, 76.30.-v, 03.67.Lx, 71.55.Cn**Abstract:**

We report Stark shift measurements for $^{121}$Sb donor electron spins in silicon using pulsed electron spin resonance. Interdigitated metal gates on top of a Sb-implanted $^{28}$Si epi-layer are used to apply electric fields. Two Stark effects are resolved: a decrease of the hyperfine coupling between electron and nuclear spins of the donor and a decrease in electron Zeeman g-factor. The hyperfine term prevails at X-band magnetic fields of 0.35T, while the g-factor term is expected to dominate at higher magnetic fields. A significant linear Stark effect is also resolved presumably arising from strain.

Since Kane's original proposal in 1998 [1], the promise of implementing quantum computation (QC) with spin qubits in silicon has generated much excitement. Advantages such as long spin decoherence times [2-4] and mature silicon technologies have been exploited and expanded in recent QC scaling strategies [5]. In donor spin QC architectures impurities are arranged in large ordered arrays on the silicon chip and placed in a strong magnetic field so that spin states of the donors can be manipulated by resonant microwave pulses. To operate on single spins in a tightly packed qubit array, electron spin Zeeman transitions might be electrostatically tuned into and out of resonance with globally applied microwaves by nanoscopic addressing gates near each donor site. The use of electric fields to shift spin resonance energies is referred to here as Stark tuning. In this Letter we lay the groundwork for these Stark tuning techniques by measuring spin resonance shifts of electrons bound to antimony donors using pulsed electron spin resonance (ESR) experiments.

In silicon the donor electron spin, $S = 1/2$, is coupled to the spin of the donor nucleus, $I \neq 0$ for all shallow donors in silicon; thus the electron spin Hamiltonian in an applied magnetic field, $B_0$, can be described to first-order by:

$$\hat{H} = g\beta B_0 S_Z + a \cdot S_Z I_Z, \qquad [1]$$

where the first term is the Zeeman electron interaction, $g$ is the electron g-factor, $\beta$ is the Bohr magneton, and the second term is the hyperfine interaction between electron and nuclear spins with hyperfine coupling constant, $a$. The $^{121}$Sb donor examined in this work has $I = 5/2$ and therefore six possible resonant Zeeman transitions for electron spin: $\nu_{M_I} = 1/h \left( g\beta B_0 + a \cdot M_I \right)$, one for each projection, $M_I$, of the $^{121}$Sb nuclear spin, accounting for six absorption lines in the ESR spectrum of $^{121}$Sb [6].

The spin resonance Stark shift of the donor electron has been theoretically studied in the context of decreasing the hyperfine interaction with the donor nuclei [7, 8]. Applying an external electric field pulls the electron wave function away from the donor nucleus, reducing the hyperfine coupling, $a$. Friesen's work [7] assumes a homogeneous electric field at each donor site which is closest to our experimental configuration. In addition to the hyperfine Stark shift, the electron g-factor may also change due to admixing with higher orbital hydrogenic states. The electric field induced g-factor change is referred to here as the spin-orbit Stark shift. By symmetry considerations, both hyperfine and spin-orbit shifts for donors in silicon should have, to lowest order, a quadratic dependence on the applied electric field. However, if the tetragonal symmetry at the donor site is broken (e.g. due to a lattice strain from random crystal defects or nearby interfaces [9, 10]), then linear Stark effects may be observed in the changes of $a$ and $g$.

Sample details were optimized for measurement with an X-band pulsed ESR spectrometer [11]. While an ideal Stark experiment would apply the same field to all donors via a parallel plate capacitor type gate structure, ESR measurements require excitation of electron spins by microwaves, which would be shielded by a parallel plate arrangement. Therefore, interdigitated metal gates were lithographically patterned on the surface of a $^{28}$Si epi-wafer implanted with $^{121}$Sb at a dose of $4\times 10^{11}$/cm$^2$ and mean implantation depth of 150nm (Fig. 1). Details of the implantation are given in reference [12]. In isotopically purified $^{28}$Si the donor electron spin decoherence is longer than in natural silicon, allowing for the long ESR pulse sequences described below. [121]Sb donors were chosen instead of [31]P in order to avoid background ESR signal from unintended phosphorus impurities in the bulk and to better control implantation depth. The metal gate lines are lengthwise and narrow to reduce absorption of the electrical microwave field in the cylindrical ESR cavity, the gate structure is a regular array across a large area to cover many donor sites for adequate ESR signal, and the gate periods are small to allow for large electric fields at small applied voltage bias. An upper limit of 3V bias between the interdigitated gates, G1 and G2, was determined by the onset of avalanche donor impact ionization current. The interdigitated gate arrangement applies a dipolar distribution of electric field strengths across the implanted donor sites, which was calculated numerically [13] (Fig. 1b).

At moderate applied electric fields the Stark shift of the donor electron spin resonance is expected to be much smaller than the inhomogeneously broadened ESR linewidths, e.g. 6.7 MHz in natural silicon due to [29]Si hyperfine interactions [14] and 0.2 MHz in isotopically-purified [28]Si due to strain [4]. Therefore this work adapts a technique developed by Mims in the early days of pulsed ESR which is sensitive to small resonance shifts by reading them out as phase shifts of the echo signal [15]. Two in-resonance microwave pulses with constant interpulse delay, $\tau$, are used to generate a two-pulse Hahn spin echo signal (Fig. 2a). In addition, electrical pulses of varied strength, polarity, and timing are applied across the interdigitated gates. Figure 2b shows six electrical pulse sequences, referred to below as experiments I-VI. If no electrical pulse is applied (experiment I), each spin precesses at a slightly different frequency due to a distribution of g-factors arising from strain, and thus during the defocusing period the spin ensemble dephases. At time $\tau$, the refocusing $\pi$-pulse reverses direction of spin precession. Since the spin precession frequency is nearly invariant throughout the experiment, all spins are in phase again at time $2\tau$ to generate an echo signal. The phase of the echo signal obtained in control experiment I is defined to be zero, and the phase of the reference microwaves are adjusted such that the entire echo signal appears in the "in-phase" channel of the microwave quadrature detector (Fig. 3a).

Via the Stark effect, electrical pulses applied during the spin echo experiment change the resonance (precession) frequencies of the electron spins. In the case of the unipolar electrical pulse (experiment II) and assuming only quadratic Stark shifts, the spin echo signal acquires an additional, uncompensated phase:

$$\varphi_{M_I} = \frac{1}{h} \cdot \left[ \Delta g(E) \cdot \beta B_0 + \Delta a(E) \cdot M_I \right] \cdot \tau \qquad [2]$$

where $\Delta g(E) = \eta_g \cdot g \cdot E^2$ and $\Delta a(E) = \eta_a \cdot a \cdot E^2$ are Stark changes to the electron g-factor and the hyperfine coupling constant, respectively, with parameters $\eta_g$ and $\eta_a$ defining the strength of the Stark effects. Because the interdigitated gate structure yields a distribution of electric fields at donor sites, there will be a distribution of Stark-induced phase shifts. As a result, averaging over the calculated distribution is required to determine the phase and magnitude of the ensemble spin echo signal. Experiment III is a control to verify that the Stark-induced phase shifts are fully refocused when a long uniform electrical pulse is applied to cover both $\tau$ periods.

In addition to the unipolar electrical pulse sequences (experiments II and III), specially designed bipolar electrical pulse sequences provide more clues to the details of the spin resonance Stark shift. In experiment IV the bipolar pulse refocuses the linear Stark effects (since their signs change with electric field polarity) and allows detection of pure quadratic Stark effects. An example is shown in Figure 3b where both a phase shift and magnitude reduction in the echo signal are observed. Alternatively, in experiment V pulses of opposite polarity and equal duration are applied during the defocusing and refocusing periods which cancel out the quadratic Stark effects and selectively detect only linear Stark shifts. Experiment VI is a differential extension of experiment IV and is useful in situations when large Stark phase shifts are desired to be observed. Since there is a broad distribution of electric fields at donor sites (Fig. 1b), the corresponding distribution of Stark-induced phase shifts fans out the electron spins and results in an echo signal intensity which is strongly diminished as compared to the echo signal at zero field. The echo signal phase shift is determined in this situation by adding the echo phase shift from the differential experiment VI to the echo phase shift due to smaller electrical pulses using experiment IV.

The main results of this work are summarized in Figure 4 by plots of echo phase shifts versus the voltage of the electrical pulses between G1 and G2. Phase shifts are measured on four hyperfine lines, $M_I = \pm 1/2$ and $\pm 5/2$, in the $^{121}$Sb spectrum allowing the estimation of individual contributions from the spin-orbit and hyperfine Stark effects. As seen in Eq. [2], phase shifts arising from the hyperfine Stark effect scale with $M_I$, whereas the spin-orbit Stark effect is equal for all lines. Thus, the observation of nearly equal and opposite phase shifts for lines $M_I = +5/2$ and -5/2 (Fig. 4b) clearly indicates that the hyperfine Stark effect dominates the phase shift of these high $M_I$ projections. The slight phase shift asymmetry observed for lines $M_I = +5/2$ and -5/2 is more pronounced for lines $M_I = +1/2$ and -1/2 (Fig. 4a), where the hyperfine shift is scaled down by a factor of 5. This asymmetry shows that the spin-orbit and hyperfine Stark shifts have the same sign and add constructively to produce a greater phase shift for $M_I = -5/2$ and -1/2, and the two effects tend to cancel each other resulting in a smaller phase shift for $M_I = +1/2$ and +5/2. From the signs of the echo phase shifts it is deduced that the hyperfine Stark effect corresponds to a decrease in hyperfine coupling, $a$, and the spin-orbit Stark effect corresponds to a decrease in the electron $g$-factor value. Fits for the four $M_I$ projections shown in Figure 4 are calculated using the same Stark hyperfine and spin-orbit parameters: $\eta_a = -3.7 \cdot 10^{-3}$ and $\eta_g = -1 \cdot 10^{-5}$ (in $\mu m^2/V^2$). Repeated measurements establish an uncertainty of 10% for hyperfine and 20% for spin-orbit parameters.

In conclusion, this study resolves two quadratic Stark effects for $^{121}$Sb donors in silicon. The measured strength of the hyperfine Stark shift, $\eta_a = -3.7 \cdot 10^{-3}$ $\mu m^2/V^2$, is smaller than that predicted for $^{31}$P donors, $\eta_a = -2 \cdot 10^{-2}$ $\mu m^2/V^2$ (as estimated from Figure 2 in [7]). No theoretical predictions are available for $^{121}$Sb donors. The spin-orbit Stark shift for donors in silicon has not previously been discussed in the literature.

Spin resonance shifts for $^{121}$Sb donor electrons are found to be small at the moderate electric fields used in this work; a maximum shift of 25 kHz was observed when average electric fields of ~0.1 V/$\mu$m were applied with $M_I = \pm 5/2$. Addressing spin qubits via resonance Stark tuning may require the use of larger electric fields nearer to the 3.7 V/$\mu$m ionization threshold [7] which is possible if the impact ionization is controlled by dilute doping and short intergate distances. Also, assuming the Stark coefficient $\eta_g$ shows no magnetic field dependence, the spin-orbit Stark effect will scale linearly with $B_0$ and will be the dominant shift mechanism at high magnetic fields, allowing for larger resonance shifts. At specific magnetic fields and nuclear spin projections ($B_0 \approx$ 1.25T for $M_I =$ +1/2, $B_0 \approx$ 3.75T for $M_I =$ +3/2, and $B_0 \approx$ 6.25T for $M_I =$ +5/2), the hyperfine and spin-orbit Stark effects should cancel each other causing the electron spin resonance frequencies to be independent of electric field.

In separate experiments using electrical pulse sequences II and V, linear Stark effects are also observed to be significant in the sample studied. In experiment V (linear Stark effects only), the electric fields caused a significant reduction in echo magnitude but no phase shift. This suggests a symmetrical distribution of intrinsic electric fields at the donor sites originating from lattice strains in the $^{28}$Si epi-layer [9]. Therefore, it is important to control strains for precise spin resonance tuning via the Stark effect. Due to a lack of quantitative information about the sample's strain distribution, this study concentrated on quantifying only the quadratic Stark terms.

This research was supported at Princeton by the Army Research Office and the Advanced Research and Development Activity under Contract No. DAAD19-02-1-0040, and at Lawrence Berkeley National Labs by NSA under ARO contract number MOD707501, the Department of Energy under Contract No. DE-AC02-05CH11231, and NSF under Grant No. 0404208. The authors would like to thank Igor Trofimov, Rogerio deSousa, and Shyam Shankar for helpful discussions.

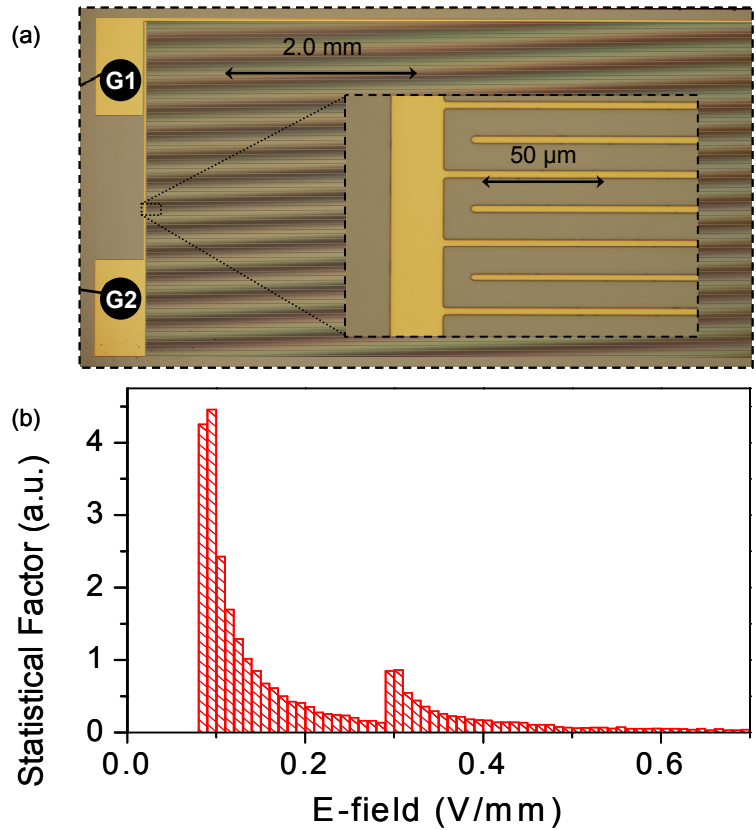

Figure 1. (Color online) (a) Micrographs of the interdigitated gate structure used in this work to measure the donor electron spin resonance Stark effect. Bonding pads for the two gates are labeled G1 and G2. The $^{28}$Si epi-layer wafer was implanted with $^{121}$Sb donors at $4*10^{11}$/cm$^2$ and a mean implantation depth ~150 nm. Gate lines are oriented along a long side of the sample (the crystallographic axis [100]) such that the microwave magnetic field in the resonator cavity is parallel to the lines and the microwave electric field is perpendicular to the lines. The device has 250 lines with 2.6μm width at 14μm period; overall dimensions of the interdigitated array were 19.5mm by 3.5mm. (b) Calculated electric field distribution at donor sites for 2 Volts applied between the interdigitated gates.

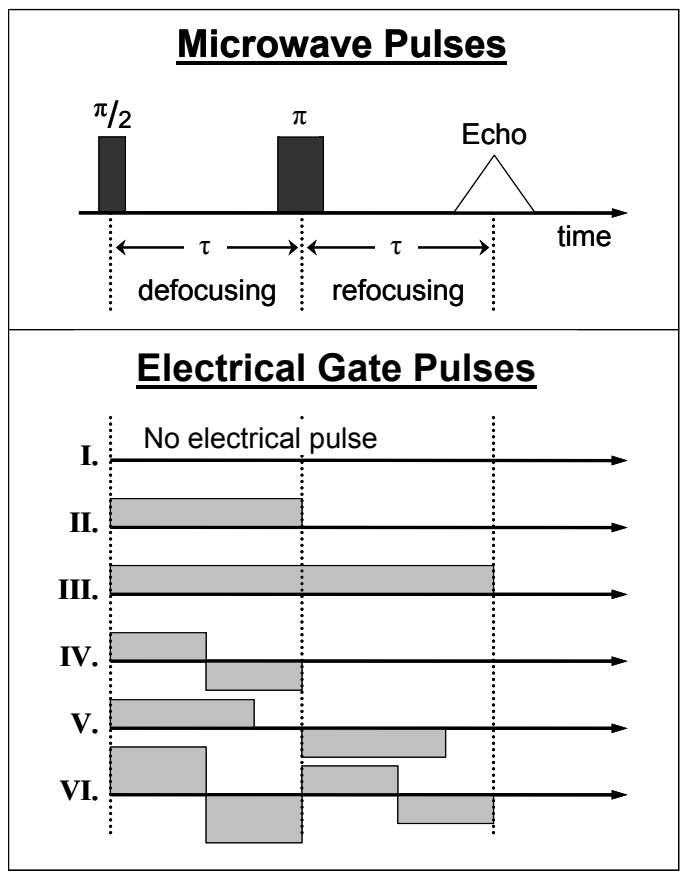

Figure 2. Pulsed ESR sequences for measuring Stark effect in the $^{28}$Si:Sb epi-layer. (a) Two microwave pulses, with rotation angles $\pi/2$ and $\pi$ and a constant interpulse delay $\tau$, generate a Hahn echo signal at time $2\tau$. (b) Voltage pulses applied between gates G1 and G2 (Fig. 1) are time correlated with the microwave pulses, such that the donor spins precess in an electric field that can be different during the defocusing and refocusing periods. The six electrical pulse sequences (I-VI) are used to discriminate the linear and quadratic Stark effects, as discussed in the text.

8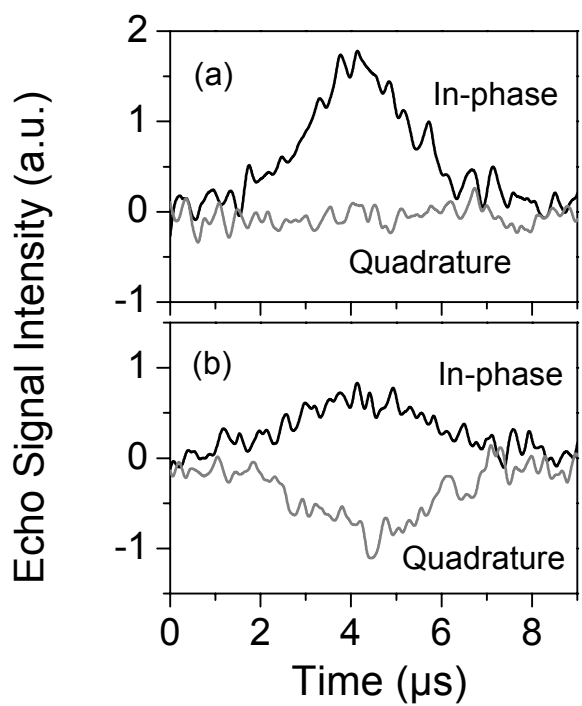

Figure 3. Traces of the two-pulse (Hahn) echo signals measured on the $M_I = +1/2$ hyperfine line for $^{28}$Si:Sb at 6.2 K using the electrical pulse sequences I (a) and IV (b) from Figure 2b. In each plot the two signals represent the in-phase and quadrature channels of the microwave quadrature detector. (a) No gate voltage is applied (sequence I) and the detector is adjusted to produce a purely in-phase echo signal. (b) A bipolar pulse, ±2V, is applied (sequence IV) during the defocusing period, and the echo signal rephases with amplitudes in both the in-phase and quadrature channel, corresponding to a phase shift of 49°.

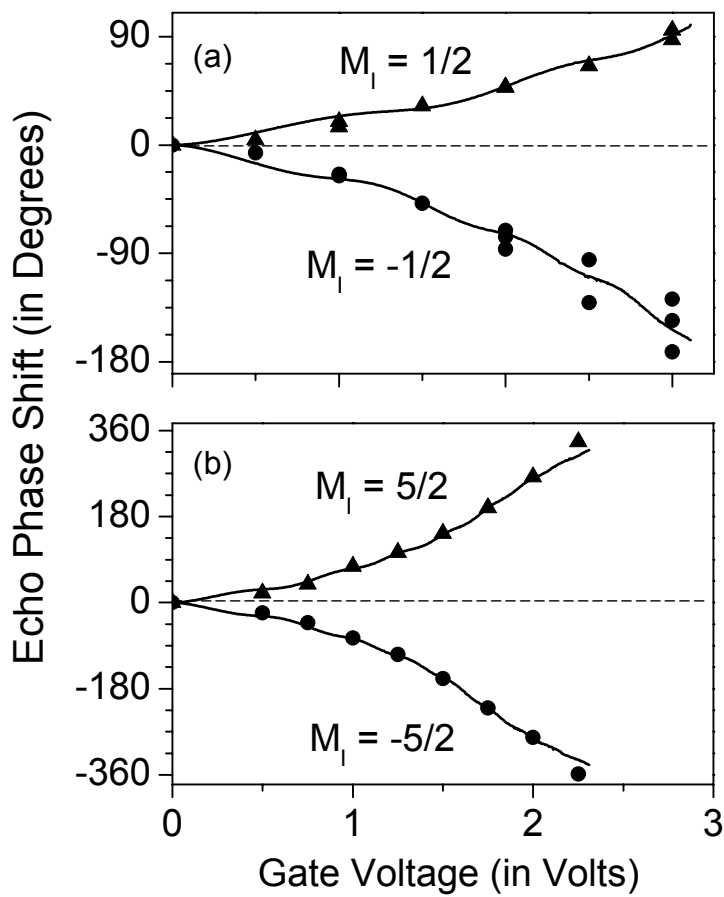

Figure 4. Phase shift of the echo signal plotted as a function of the applied gate voltage for $^{28}$Si:Sb at 6.2 K and the interpulse delay, $\tau = 40\mu s$. Bipolar electrical pulse sequences IV and VI (Fig. 2b) were used and therefore the phase shifts are due to quadratic Stark effects only. Data (points) and numerical fits (lines) are shown for the $M_I = \pm 1/2$ (a) and $M_I = \pm 5/2$ (b) hyperfine lines of the six-line ESR spectrum of $^{121}$Sb. Note that phase shifts for the $M_I = \pm 5/2$ lines are extended beyond 1.5V with the differential voltage sequence (experiment VI). For the $M_I = -1/2$ line, the data in the 2-3V range were obtained using both regular (IV) and differential (VI) voltage sequences. Errors in determining echo phase shifts were larger for larger shifts, but were generally less than $\pm 5°$. All four $M_I$ lines were fit using the same pair of hyperfine and spin-orbit Stark parameters.